\documentclass[12pt]{iopart}
\usepackage{graphicx}
\eqnobysec

\newcommand{\beq}{\begin{equation}}
\newcommand{\beqa}{\begin{eqnarray}}
\newcommand{\eeq}{\end{equation}}
\newcommand{\eeqa}{\end{eqnarray}}
\newcommand{\al}{\alpha}
\newcommand{\be}{\beta}

\renewcommand{\d}{{\rm d}}

\newcommand{\ea}{{\rm e}^{2J_A}}
\newcommand{\eb}{{\rm e}^{2J_B}}
\newcommand{\ec}{{\rm e}^{2J_C}}
\newcommand{\eab}{{\rm e}^{2J_A+2J_B}}
\newcommand{\eac}{{\rm e}^{2J_A+2J_C}}
\newcommand{\ebc}{{\rm e}^{2J_B+2J_C}}

\newcommand{\eps}{\varepsilon}
\newcommand{\frad}[2]{\displaystyle{\displaystyle#1\over\displaystyle#2}}
\newcommand{\haut}{\rule[-14pt]{0pt}{36pt}}
\renewcommand{\in}{{\rm in}}
\newcommand{\lef}{{\rm left}}
\newcommand{\mean}[1]{\left\langle#1\right\rangle}
\newcommand{\out}{{\rm out}}
\newcommand{\rig}{{\rm right}}
\newcommand{\s}{{\rm st}}
\renewcommand{\u}{u}
\newcommand{\w}{w}
\newcommand{\x}{x}
\newcommand{\y}{y}
\newcommand{\C}{{\cal C}}
\newcommand{\D}{\Delta}
\renewcommand{\H}{{\cal H}}
\newcommand{\K}{{\rm K}}
\renewcommand{\L}{{\rm L}}
\newcommand{\M}{{\rm M}}
\newcommand{\N}{{\rm N}}

\begin{document}

\title[Nonequilibrium stationary states with Gibbs measure
for two or three species]
{Nonequilibrium stationary states with Gibbs measure
for two or three species of interacting particles}

\author{J M Luck\dag\ and C Godr\`{e}che\ddag}

\address{\dag\ Service de Physique Th\'eorique\footnote{URA 2306 of CNRS},
CEA Saclay, 91191 Gif-sur-Yvette cedex, France}

\address{\ddag\ Isaac Newton Institute for Mathematical Sciences,
20 Clarkson Road, Cambridge\break CB3 0EH, UK, and
Service de Physique de l'\'Etat Condens\'e\footnote{URA 2464 of CNRS},
CEA Saclay, 91191 Gif-sur-Yvette cedex, France}

\begin{abstract}
We construct explicit examples of one-dimensional driven diffusive sys\-tems
for two and three species of interacting particles,
defined by asymmetric dynamical rules which do not obey detailed balance,
but whose nonequilibrium stationary-state measure coincides
with a prescribed equilibrium Gibbs measure.
For simplicity, the measures considered in this construction
only involve nearest-neighbor interactions.
For two species,
the dynamics thus obtained generically has five free parameters,
and does not obey pairwise balance in general.
The latter property is satisfied only by the totally asymmetric dynamics
and the partially asymmetric dynamics with uniform bias,
i.e., the cases originally considered by Katz, Lebowitz, and Spohn.
For three species of interacting particles,
with nearest-neighbor interactions between particles of the same species,
the totally asymmetric dynamics thus obtained has two free parameters,
and obeys pairwise balance.
These models are put in perspective with other examples of driven
diffusive systems.
The emerging picture is that asymmetric (nonequilibrium) stochastic dynamics
leading to a given stationary-state measure
are far more constrained (in terms of numbers of free parameters)
than the corresponding symmetric (equilibrium) dynamics.
\end{abstract}


\eads{
\mailto{luck@dsm-mail.saclay.cea.fr},
\mailto{godreche@dsm-mail.saclay.cea.fr}}

\maketitle

\section{Introduction}

Driven diffusive systems~\cite{s,sz,l,sch}
are defined by stochastic dynamical rules
that incorporate the effect of an external drive,
and therefore do not obey detailed balance, which makes
their nonequilibrium stationary states difficult to study in general.
The first step of evaluating the stationary-state measure,
i.e., the probability $P_\s(\C)$
of any given configuration~$\C$, is already a non-trivial task.
In view of the lack of a general theory of nonequilibrium stationary states,
one has to rely on the investigation of specific models,
for which the stationary-state measure is analytically tractable.
Examples of such models are the simple exclusion processes,
the zero-range process, and the one-dimensional Katz-Lebowitz-Spohn (KLS)
model.

For the simple exclusion process,
particles obeying an exclusion constraint perform random walks on a lattice,
with either symmetric or biased moves~\cite{l,sch,der}.
On a ring, i.e., with periodic boundary conditions,
the stationary-state measure is uniform, irrespectively of the bias:
all the configurations are equally probable.
The same model with open boundaries has a stationary-state measure
which can be described in terms of a matrix product Ansatz.
The zero-range process (ZRP)~\cite{spitzer}, in the presence of a bias,
belongs to the class of driven stochastic
processes with multiple occupancies.
Its stationary state has a product measure, again irrespectively of the bias.
The occupation numbers of the sites are independent random
quantities with a common distribution,
up to the conservation of the total number of particles.
The existence of unbounded occupancies however opens up the possibility
of having a condensation transition,
irrespective of the geometry of the underlying lattice,
and therefore also in one dimension~\cite{ev,lux}.

The KLS model is a lattice gas model of interacting charged particles
subjected to an external electric field~\cite{kls}.
It is representative of the class of models with non-equilibrium stationary
state measures incorporating physical interactions between particles.
The stationary-state measure of the KLS model is non trivial
in two dimensions and above, where e.g.~the critical temperature
depends continuously on the applied field~\cite{sz}.
The situation however simplifies in the one-dimensional geometry,
where the model is equivalent to a chain of classical Ising spins $s_n=\pm1$.
In this case, there exists a class of biased stochastic dynamics,
for which the nonequilibrium stationary-state measure
is the canonical finite-temperature Gibbs measure,
where the probability $P_\s(\C)$ of the configuration $\C=\{s_1,\dots,s_N\}$
is given by the Boltzmann formula (with $k_BT=1$)
\beq
P_\s(\C)=\frac{1}{Z}\,\exp(-\H(\C))
\label{boltz}
\eeq
associated with the Ising Hamiltonian with nearest-neighbor interactions
\beq
\H=-J\sum_ns_ns_{n+1}.
\label{ham2}
\eeq
This very stochastic model with antiferromagnetic interactions ($J<0$)
was subsequently
rediscovered in the context of polymer crystallization~\cite{gw}.

At this point it is natural to question the generality of the result
found in~\cite{kls} for the case of the Ising chain.
In the present work we show that for systems with three species of particles
there also exist asymmetric stochastic dynamics
which do not obey detailed balance,
but whose stationary-state measure is the Gibbs measure
corresponding to a simple Hamiltonian.
In particular this stationary-state measure is independent of the bias.
In our construction we restrict the choice of measures
to those involving only nearest-neighbor interactions.
We first revisit in Section~2 the case of two species of interacting particles,
thus generalizing the study of~\cite{kls} to a wider class of dynamics.
The emphasis is put on the role of various symmetries,
and especially on the number of free parameters left by imposing them.
We then consider, in Section~3, the entirely novel situation
of three species of interacting particles.
Section~4 contains a discussion, where our results are put in perspective
with yet other examples.

Let us finally give a brief reminder of the concepts
of detailed balance~\cite{kampen} and pairwise balance~\cite{pairw},
which will be used throughout this work.
Consider a finite set of configurations $\C$,
and a Markovian dynamics in continuous time,
defined by the transition rates $W(\C\to\C')$.
The master equation for the time-dependent probability $P(\C,t)$ reads
\beq
\frac{\d P(\C,t)}{\d t}
=\sum_{\C'}\Bigl(W(\C'\to\C)P(\C',t)-W(\C\to\C')P(\C,t)\Bigr).
\eeq
The stationary probability $P_\s(\C)$ therefore obeys the equation
\beq
\sum_{\C'}\Bigl(W(\C'\to\C)P_\s(\C')-W(\C\to\C')P_\s(\C)\Bigr)=0.
\label{bal}
\eeq

In the particular case where the Markov process is reversible,
the dynamics brings the system to an equilibrium state.
Reversibility requires the {\it detailed balance} property~\cite{kampen},
that is the absence of probability flux
between any pair of configurations~$\C$ and $\C'$ at stationarity:
\beq
W(\C\to\C')P_\s(\C)=W(\C'\to\C)P_\s(\C').
\label{detbal}
\eeq
Equation~(\ref{detbal}) clearly implies~(\ref{bal}).

A weaker property, referred to as {\it pairwise balance}~\cite{pairw},
is adapted to the situation of driven diffusive systems,
where the presence of a preferred direction of motion, i.e., a bias,
precludes the property of detailed balance.
Pairwise balance is defined as follows:
for every pair of configurations $\C$ and $\C'$ such that $W(\C\to\C')\neq0$,
there exists a third configuration $\C''$ such that
\beq
W(\C\to\C')P_\s(\C)=W(\C''\to\C)P_\s(\C'').
\label{pairbal}
\eeq
The moves $\C\to\C'$ and $\C''\to\C$ are said to be conjugate to each other.
Equation~(\ref{pairbal}) also implies~(\ref{bal}),
since terms corresponding to pairs of conjugate moves cancel each other.
Several of the examples of dynamics constructed in this paper
obey P-related pairwise balance,
where pairs of conjugate moves are related to each other by parity.
The first explicit example is given below~(\ref{eq3}).
An important general consequence of P-related pairwise balance
is shown at the end of Section~3.
If the totally asymmetric dynamics obeys P-related pairwise balance,
the partially asymmetric dynamics
with uniform bias~$p$ also obeys pairwise balance,
and it has the same stationary-state measure as the totally asymmetric one,
irrespective of the value of the bias.

\section{Two species}

Consider a ring made of $N$ sites.
Each site is occupied by a particle,
which can be either of type $A$ (positively charged)
or of type $B$ (negatively charged).
We represent the species of particle at site $n$
by an Ising spin $s_n=\pm1$ equal to the charge of the particle.
Table~\ref{2s} also gives the corresponding indicator variables.
For instance, $(1+s_n)/2$ is equal to 1 if the particle at site $n$
is of type $A$, and to 0 else.

\begin{table}[ht]
\caption{Spin (charge) and indicator variables
associated with each particle species in the case of two species.}
\label{2s}
\begin{center}
\begin{tabular}{|c|c|c|}
\hline
Species at site $n$&Spin (charge)&Indicator variable\\
\hline
$A$&$s_n=+1$&$(1+s_n)/2$\\
$B$&$s_n=-1$&$(1-s_n)/2$\\
\hline
\end{tabular}
\end{center}
\end{table}

Our goal is to construct nonequilibrium dynamics
such that the stationary-state measure is the Gibbs measure
given by~(\ref{boltz}) associated with
the nearest-neighbor Hamiltonian~(\ref{ham2}).
We consider the asymmetric exchange (Kawasaki) dynamics.
Consistently with the form of the Hamiltonian~(\ref{ham2}),
the rates depend on the two neighbors
of the pair to be exchanged, according to Table~\ref{2d}.
The dynamics so defined conserves the numbers $N_A$ and $N_B$
of particles of each species, with $N_A+N_B=N$.
We first determine the most general dynamics of this form
in Section~2.1, and then discuss the interplay between various possible
symmetries in Section~2.2.

\begin{table}[ht]
\caption{List of moves in the general exchange dynamics
for two species of interacting particles,
notation for the corresponding exchange rates,
and energy difference $\D\H$ involved in the moves,
where the Hamiltonian $\H$ is defined in~(\ref{ham2}).}
\label{2d}
\begin{center}
\begin{tabular}{|c|c|c||c|c|c|}
\hline
Move&Rate&$\D\H$&Move&Rate&$\D\H$\\
\hline
$AABA\to ABAA$&$\w_{AA}$&$0$&
$ABAA\to AABA$&$\x_{AA}$&$0$\\
$AABB\to ABAB$&$\w_{AB}$&$4J$&
$ABAB\to AABB$&$\x_{AB}$&$-4J$\\
$BABA\to BBAA$&$\w_{BA}$&$-4J$&
$BBAA\to BABA$&$\x_{BA}$&$4J$\\
$BABB\to BBAB$&$\w_{BB}$&$0$&
$BBAB\to BABB$&$\x_{BB}$&$0$\\
\hline
\end{tabular}
\end{center}
\end{table}

\subsection{The general case}

Consider the numbers $N_{AA},\dots,N_{BB}$ of oriented pairs
of neighbors of each species.
These numbers obey the sum rules
\beq
\fl N_A=N_{AA}+N_{AB}=N_{AA}+N_{BA},\qquad N_B=N_{BA}+N_{BB}=N_{AB}+N_{BB}.
\label{paireqs}
\eeq
The sum of the two equations gives twice the same equation,
so that the four pair numbers obey three independent equations,
leaving one single free quantity.
It is convenient to take the latter as being the Hamiltonian $\H$
of~(\ref{ham2}).
The pair numbers can indeed be expressed
as linear combinations of $\H$ and of the particle numbers $N_A$ and~$N_B$:
\beq
\matrix{
N_{AA}=\frad{1}{4}\left(3N_A-N_B-\H/J\right),\quad\hfill
&N_{BB}=\frad{1}{4}\left(3N_B-N_A-\H/J\right),\cr
N_{AB}=N_{BA}=\frad{1}{4}\left(N+\H/J\right).\hfill&
}
\label{pairex}
\eeq
Incidentally, this proves that the Hamiltonian~(\ref{ham2})
is the most general form of a pair Hamiltonian for two species of particles
with nearest-neighbor interactions.

Throughout this paper, we make use of an alternative and convenient way
of automatically taking into account sum rules such as~(\ref{paireqs}).
This consists in expressing all the quantities
in terms of the spin variables $s_n$ introduced in Table~\ref{2s}.
For instance
\beq
N_A=\frac{1}{2}\sum_n(1+s_n)=\frac{N}{2}\,(1+\mean{s_1}).
\label{cordef}
\eeq
Here and in the following, the brackets $\mean{\dots}$ denote
a uniform spatial average {\it for a fixed generic configuration~$\C$.}
Recall that all the sites are equivalent, because of translational invariance.
The pair numbers and the Hamiltonian read
\beq
\matrix{
\haut N_{AA}=\frad{N}{4}\,(1+2\mean{s_1}+\mean{s_1s_2}),\quad\hfill
&N_{BB}=\frad{N}{4}\,(1-2\mean{s_1}+\mean{s_1s_2}),\hfill\cr
\haut N_{AB}=N_{BA}=\frad{N}{4}\,(1-\mean{s_1s_2}),\quad\hfill
&\H=-NJ\mean{s_1s_2}.\hfill
}
\eeq
Equations such as~(\ref{paireqs}) and~(\ref{pairex})
are then automatically satisfied.

Consider now the fate of a generic configuration $\C$.
The total exit rate $W_\out(\C)$ from~$\C$ to any other configuration $\C'$
per unit time can be read off from Table~\ref{2d}:
\beqa
&&W_\out(\C)
=\w_{AA}N_{AABA}+\w_{AB}N_{AABB}+\w_{BA}N_{BABA}+\w_{BB}N_{BABB}\nonumber\\
&&{\hskip 14.3truemm}
+\x_{AA}N_{ABAA}+\x_{AB}N_{ABAB}+\x_{BA}N_{BBAA}+\x_{BB}N_{BBAB}.
\label{wout}
\eeqa
An analogous expression can be derived for the total entrance rate
$W_\in(\C)$ from any other configuration $\C'$ to $\C$.
Using again Table~\ref{2d}, as well as~(\ref{boltz}) to express
the stationary-state weight $P_\s(\C')$ as
\beq
P_\s(\C')=P_\s(\C)\exp(\D\H),
\label{delham}
\eeq
in terms of $P_\s(\C)$ and of the energy difference
\beq
\D\H=\H(\C)-\H(\C')
\eeq
involved in the move, we obtain
\beqa
&&\fl W_\in(\C)=
\w_{AA}N_{ABAA}+{\rm e}^{4J}\w_{AB}N_{ABAB}+{\rm e}^{-4J}\w_{BA}N_{BBAA}+\w_{BB}N_{BBAB}
\nonumber\\
&&\fl{\hskip 12.7truemm}
+\x_{AA}N_{AABA}+{\rm e}^{-4J}\x_{AB}N_{AABB}+{\rm e}^{4J}\x_{BA}N_{BABA}+\x_{BB}N_{BABB}.
\label{win}
\eeqa
In the stationary state we have
\beq
W_\out(\C)-W_\in(\C)=0
\label{winwout}
\eeq
for every configuration~$\C$.
In order to determine the number of independent conditions on the rates
imposed by this equation,
it is convenient to rewrite~(\ref{wout}) and~(\ref{win})
in terms of spin correlations, i.e., spatial averages of products of spin
variables, denoted as~$\mean{\dots}$, along the lines of~(\ref{cordef}).
With these notations, we obtain
\beqa
&&\fl W_\out(\C)-W_\in(\C)=\frac{N}{16}\,\bigg\{
({\rm e}^{-4J}-1)(\mean{s_1s_2s_3s_4}+1)\,R_1
+\mean{s_1(s_3-s_2)s_4}\,R_2\nonumber\\
&&\fl{\hskip 42truemm}
+\Big[\mean{s_1(3s_2-2s_3+s_4)}+{\rm e}^{-4J}\mean{s_1(s_2-2s_3-s_4)}\Big]
\,R_1\bigg\},
\label{dw}
\eeqa
where $R_1$ and $R_2$ stand for the following linear combinations of the rates:
\beqa
&&R_1={\rm e}^{4J}(\w_{AB}+\x_{BA})-\w_{BA}-\x_{AB},\nonumber\\
&&R_2=({\rm e}^{4J}+1)(\w_{AB}-\x_{BA})+({\rm e}^{-4J}+1)(\w_{BA}-\x_{AB})
\nonumber\\
&&{\hskip 5truemm}-2(\w_{AA}+\w_{BB}-\x_{AA}-\x_{BB}).
\label{r1r2}
\eeqa
The condition~(\ref{winwout}) therefore gives two linear relations,
\beq
R_1=R_2=0,
\label{eq}
\eeq
between the eight exchange rates defining the general asymmetric dynamics.
Let us choose the time unit by setting
\beq
\w_{AA}+\w_{BB}+\x_{AA}+\x_{BB}=1.
\label{2nor}
\eeq

The most general asymmetric dynamics for two species of interacting particles
such that the stationary-state measure is given by~(\ref{boltz}),~(\ref{ham2})
therefore has five free parameters.
An explicit parametrization of the rates is given in Table~\ref{2gene}.
The dynamics thus obtained does not obey pairwise balance in general.

The parametrization of the solutions to~(\ref{eq}) and~(\ref{2nor})
given in Table~\ref{2gene} has been carefully chosen
in such a way that the various symmetries
to be described below correspond to the simple
constraints~(\ref{cstr1}), (\ref{cstr2}), (\ref{cstr3}), (\ref{cstr4})
in terms of the parameters~$\delta$ and $\eps_1,\dots,\eps_4$.
The parameters~$\delta$ and $\eps_1,\dots,\eps_4$
all lie in the range $[-1,+1]$, and are such that the combination
\beq
\lambda=\frac{(1+\delta)\eps_1+(1-\delta)\eps_2}{\eps_3+\eps_4}
\label{lamdef}
\eeq
is positive.

\begin{table}[ht]
\caption{Explicit parametrization of the rates
of the most general asymmetric dynamics for two species
with Gibbs stationary-state measure~(\ref{boltz}),~(\ref{ham2}).
The notation $\lambda$ is defined in~(\ref{lamdef}).}
\label{2gene}
\begin{center}
\begin{tabular}{|c|c||c|c|}
\hline
Rate&expression&Rate&expression\\
\hline
\haut$\w_{AA}$&$\frad{(1+\eps_1)(1+\delta)}{4}$&
\haut$\x_{AA}$&$\frad{(1-\eps_1)(1+\delta)}{4}$\\
\haut$\w_{AB}$&$\frad{(1+\eps_3)\lambda}{2({\rm e}^{4J}+1)}$&
\haut$\x_{AB}$&$\frad{(1-\eps_4)\lambda\,{\rm e}^{4J}}{2({\rm e}^{4J}+1)}$\\
\haut$\w_{BA}$&$\frad{(1+\eps_4)\lambda\,{\rm e}^{4J}}{2({\rm e}^{4J}+1)}$&
\haut$\x_{BA}$&$\frad{(1-\eps_3)\lambda}{2({\rm e}^{4J}+1)}$\\
\haut$\w_{BB}$&$\frad{(1+\eps_2)(1-\delta)}{4}$&
\haut$\x_{BB}$&$\frad{(1-\eps_2)(1-\delta)}{4}$\\
\hline
\end{tabular}
\end{center}
\end{table}

\subsection{The interplay between various symmetries}

The number of free parameters of the dynamics thus obtained
is decreased if various kinds of symmetries are imposed onto the dynamics.

\medskip\noindent$\bullet$
{\it Symmetric ({\rm P}-invariant) dynamics.}
Consider a symmetric dynamics, invariant under the spatial parity P
which reverses the orientation of the ring (i.e., interchanges left and right).
This symmetry property reads
\beq
\x_{IJ}=\w_{JI}
\label{eq1}
\eeq
for all values of the indices $I,J=A,B$.
The stationary state thus obtained is an equilibrium state.
The first equation of~(\ref{eq}),
\beq
\w_{BA}={\rm e}^{4J}\w_{AB},
\label{deb}
\eeq
expresses detailed balance.
Equation~(\ref{eq1}) amounts to setting
\beq
\eps_i=0\quad(i=1,\dots,4)
\label{cstr1}
\eeq
in Table~\ref{2gene}.
The symmetric (equilibrium) dynamics therefore has two free parameters:
$\delta$ and $\lambda$, in the ranges $-1<\delta<1$ and $\lambda>0$.
The expression~(\ref{lamdef}) for $\lambda$ indeed becomes indeterminate
in the limit where all the $\eps_i$ go simultaneously to zero.
The rates read
\beq
\matrix{
\haut\w_{AA}=\x_{AA}=\frad{1+\delta}{4},\quad\hfill
&\w_{AB}=\x_{BA}=\frad{\lambda}{2({\rm e}^{4J}+1)},\cr\hfill
\haut\w_{BA}=\x_{AB}=\frad{\lambda\,{\rm e}^{4J}}{2({\rm e}^{4J}+1)},\quad\hfill
&\w_{BB}=\x_{BB}=\frad{1-\delta}{4}.\hfill
}
\label{ress}
\eeq

\medskip\noindent$\bullet$
{\it Totally asymmetric dynamics.}
Consider a dynamics driven by an infinitely strong electric field,
so that the positively (resp.~negatively) charged $A$ particles
(resp.~$B$ particles) hop exclusively to the right (resp.~to the left).
Therefore
\beq
\x_{IJ}=0
\label{eq2}
\eeq
for all values of the indices $I,J=A,B$.
Equation~(\ref{eq}) becomes
\beq
\w_{BA}={\rm e}^{4J}\w_{AB}.
\label{eq3}
\eeq
This equation coincides with~(\ref{deb}).
It expresses {\it {\rm P}-related pairwise balance:}
conjugate moves are related to each other by parity P,
i.e., the first and the fourth move of the left column of Table~\ref{2d}
are their own conjugates,
whereas the second and the third moves are conjugate to each other.
Equation~(\ref{eq2}) amounts to setting
\beq
\eps_i=1\quad(i=1,\dots,4)
\label{cstr2}
\eeq
in Table~\ref{2gene}, so that $\lambda=1$.
The totally asymmetric dynamics
with stationary-state measure~(\ref{boltz}),~(\ref{ham2})
therefore has one free parameter:
$\delta$, in the range $-1<\delta<1$.
The rates read
\beq
\matrix{
\haut\w_{AA}=\frad{1+\delta}{2},\quad\hfill
&\w_{AB}=\frad{1}{{\rm e}^{4J}+1},\hfill\cr
\w_{BA}=\frad{{\rm e}^{4J}}{{\rm e}^{4J}+1},\quad\hfill
&\w_{BB}=\frad{1-\delta}{2}.\hfill
}
\label{resas}
\eeq

\medskip\noindent$\bullet$
{\it Partially asymmetric dynamics with a uniform bias.}
This is the most general case originally considered by KLS~\cite{kls}.
Consider a dynamics driven by a finite electric field,
so that the positively (resp.~negatively) charged $A$ particles
(resp.~$B$ particles) hop preferentially to the right (resp.~to the left).
Let
\beq
p=\frad{1+\eps}{2},\qquad q=\frad{1-\eps}{2}
\eeq
be the a priori probabilities of respectively hopping
to the right and to the left,
where $0\le\eps\le1$ provides a measure of the applied electric field.
This translates into the following uniform bias condition:
\beq
\frac{\x_{IJ}}{\w_{JI}}=\frac{1-\eps}{1+\eps}
\label{eqpq}
\eeq
for all values of $I,J=A,B$.
This situation interpolates between the symmetric case ($p=1/2$, $\eps=0$)
and the totally asymmetric one ($p=1$, $\eps=1$).
Equation~(\ref{eqpq}) amounts to setting
\beq
\eps_i=\eps\quad(i=1,\dots,4)
\label{cstr3}
\eeq
in Table~\ref{2gene}, so that again $\lambda=1$.
As a consequence, there is a two-parameter family of dynamics
with uniform bias and stationary-state
measure~(\ref{boltz}),~(\ref{ham2}), parametrized by~$\eps$ and $\delta$.
We thus recover the original KLS model~\cite{kls}.
The fact that the stationary-state weights are independent of the bias
is actually a general property of dynamics obeying P-related pairwise balance
(see Section~3.2).

\medskip\noindent$\bullet$
{\it {\rm CP}-invariance.}
The CP operation is the product of C and P,
where the charge conjugation C changes the charge of the particles
to its opposite (i.e., interchanges $A$ and~$B$ particles),
whereas the spatial parity P
changes the orientation of the ring (i.e., interchanges left and right).
In physical terms, in the stationary state of a CP-invariant dynamics,
the current due to a positively charged particle
and to a negatively charged particle are equal.
Requiring CP-invariance yields the two conditions
\beq
\w_{AA}=\w_{BB},\qquad\x_{AA}=\x_{BB},
\label{eqcp}
\eeq
which amount to setting
\beq
\delta=0,\qquad\eps_1=\eps_2=\eps
\label{cstr4}
\eeq
in Table~\ref{2gene}.
The most general CP-invariant dynamics
has therefore three free pa\-ra\-me\-ters:~$\eps$, $\eps_3$, and $\eps_4$.
It does not obey pairwise balance in general.

CP-invariance can be combined with any of the above symmetries:

\smallskip\noindent $\star$
The CP-invariant symmetric dynamics corresponds to $\delta=\eps_i=0$.
It has a single free parameter: $\lambda$.
The rates read
\beq
\matrix{
\haut\w_{AA}=\x_{AA}=\frad{1}{4},\hfill\quad
&\w_{AB}=\x_{BA}=\frad{\lambda}{2({\rm e}^{4J}+1)},\hfill\cr
\haut\w_{BA}=\x_{AB}=\frad{\lambda\,{\rm e}^{4J}}{2({\rm e}^{4J}+1)},\hfill\quad
&\w_{BB}=\x_{BB}=\frad{1}{4}.\hfill
}
\eeq

\smallskip\noindent $\star$
The partially asymmetric CP-invariant dynamics with uniform bias
has one single free parameter: $\eps$.

\smallskip\noindent $\star$
The totally asymmetric CP-invariant dynamics
is the most constrained of all the dynamics: it has no free parameter at all.
The rates
\beq
\w_{AA}=\w_{BB}=\frac{1}{2},\qquad\w_{AB}=\frac{1}{{\rm e}^{4J}+1},\qquad
\w_{BA}=\frac{{\rm e}^{4J}}{{\rm e}^{4J}+1}
\label{resascp}
\eeq
only depend on the energy difference $\D\H$ involved in the exchange moves.
They coincide with those of the heat-bath rule~\cite{binder,krauth}:
\beq
w(\D\H)=\frac{1}{\exp(\D\H)+1}
=\frac{1}{2}\left(1-\tanh\frac{\D\H}{2}\right).
\label{hb}
\eeq

The above discussion is summarized in Table~\ref{2summary},
giving the number of free parameters
for every symmetry class of dynamics, both without and with imposing
CP-invariance.

\begin{table}[ht]
\caption{
List of the symmetry classes of dynamics for two species of particles
with Gibbs stationary-state measure~(\ref{boltz}),~(\ref{ham2}),
with balance property: detailed balance (DB) or pairwise balance (PB),
and number of free parameters, both without and with CP-invariance.}
\label{2summary}
\begin{center}
\begin{tabular}{|l|c|c|c|}
\hline
Class of dynamics&Balance property&Without CP&With CP\\
\hline
General&none&5&3\\
Symmetric (equilibrium)&DB&2&1\\
Totally asymmetric&PB&1&0\\
Partially asymmetric (uniform bias)&PB&2&1\\
\hline
\end{tabular}
\end{center}
\end{table}

\section{Three species}

Consider again a finite ring of $N$ sites.
Each site is now occupied by a particle which can be
either of type $A$ (positively charged), of type $B$ (negatively charged),
or of type~$C$ (neutral, i.e., with no charge).
We again represent the species of particle at site $n$
by a spin $S_n=0,\pm1$ equal to the charge of the particle,
as shown in Table~\ref{3s}.

\begin{table}[ht]
\caption{Spin (charge) and indicator variables
associated with each particle species in the case of three species.}
\label{3s}
\begin{center}
\begin{tabular}{|c|c|c|}
\hline
Species at site $n$&Spin (charge)&Indicator variable\\
\hline
$A$&$S_n=+1$&$S_n(S_n+1)/2$\\
$B$&$S_n=-1$&$S_n(S_n-1)/2$\\
$C$&$S_n=0$&$1-S_n^2$\\
\hline
\end{tabular}
\end{center}
\end{table}

We consider Gibbs measures corresponding to the most general
(ferromagnetic or antiferromagnetic) Hamiltonian involving pairs of identical
nearest neighbors:
\beq
\H=-2(J_AN_{AA}+J_BN_{BB}+J_CN_{CC}),
\label{ham3}
\eeq
where the coupling constants $J_A$, $J_B$, and $J_C$ can take both signs.
The factor 2 is introduced for consistency with the case of two species.
Using the spin variables $S_n=0,\pm1$ defined in Table~\ref{3s},
the Hamiltonian~(\ref{ham3}) can be rewritten as
\beq
\fl\H=E_0-\case{1}{2}\sum_n
\Bigl[(J_A+J_B+4J_C)S_nS_{n+1}+(J_A-J_B)(S_n+S_{n+1})+J_A+J_B\Bigr]
S_nS_{n+1},\;
\label{hambeg}
\eeq
where $E_0=2(N-2N_C)J_C$ is a constant.
This is a generalized Blume-Emery-Griffiths spin-1 Hamiltonian~\cite{beg}.

We again address the question of the existence of nonequilibrium stochastic
dynamics whose stationary-state measure is the measure~(\ref{boltz})
associated with the Hamiltonian~(\ref{ham3}).
The results obtained in Section~2 for two species of particles suggest
that the case of totally asymmetric exchange dynamics is already of interest.
We therefore restrict our investigation to this limiting situation
for the time being.
The positively (resp.~negatively) charged~$A$ particles
(resp.~$B$ particles) only hop to the right (resp.~to the left),
whereas the neutral~$C$ particles can hop in both directions.
The exchange rates depend on the two neighbors
of the pair to be exchanged, according to Table~\ref{3d}.
The dynamics so defined conserves the numbers $N_A$, $N_B$, and $N_C$
of particles of each species, with $N_A+N_B+N_C=N$.

\begin{table}[ht]
\caption{List of moves in the totally asymmetric dynamics
for three species of interacting particles,
notation for the corresponding exchange rates,
and energy difference $\D\H$ involved in the moves,
where the Hamiltonian $\H$ is defined in~(\ref{ham3}).}
\label{3d}
\begin{center}
\begin{tabular}{|l|c|c||c|c|c|}
\hline
Move&Rate&$\D\H$&Move&Rate&$\D\H$\\
\hline
$AABA\to ABAA$&$\w_{AA}$&$0$&
$BCBC\to BBCC$&$\x_{BC}$&$-2J_B-2J_C$\\
$AABB\to ABAB$&$\w_{AB}$&$2J_A+2J_B$&
$CCBA\to CBCA$&$\x_{CA}$&$2J_C$\\
$AABC\to ABAC$&$\w_{AC}$&$2J_A$&
$CCBB\to CBCB$&$\x_{CB}$&$2J_B+2J_C$\\
$BABA\to BBAA$&$\w_{BA}$&$-2J_A-2J_B$&
$CCBC\to CBCC$&$\x_{CC}$&$0$\\
$BABB\to BBAB$&$\w_{BB}$&$0$&
$AACA\to ACAA$&$\y_{AA}$&$0$\\
$BABC\to BBAC$&$\w_{BC}$&$-2J_B$&
$AACB\to ACAB$&$\y_{AB}$&$2J_A$\\
$CABA\to CBAA$&$\w_{CA}$&$-2J_A$&
$AACC\to ACAC$&$\y_{AC}$&$2J_A+2J_C$\\
$CABB\to CBAB$&$\w_{CB}$&$2J_B$&
$BACA\to BCAA$&$\y_{BA}$&$-2J_A$\\
$CABC\to CBAC$&$\w_{CC}$&$0$&
$BACB\to BCAB$&$\y_{BB}$&$0$\\
$ACBA\to ABCA$&$\x_{AA}$&$0$&
$BACC\to BCAC$&$\y_{BC}$&$2J_C$\\
$ACBB\to ABCB$&$\x_{AB}$&$2J_B$&
$CACA\to CCAA$&$\y_{CA}$&$-2J_A-2J_C$\\
$ACBC\to ABCC$&$\x_{AC}$&$-2J_C$&
$CACB\to CCAB$&$\y_{CB}$&$-2J_C$\\
$BCBA\to BBCA$&$\x_{BA}$&$-2J_B$&
$CACC\to CCAC$&$\y_{CC}$&$0$\\
$BCBB\to BBCB$&$\x_{BB}$&$0$&&&\\
\hline
\end{tabular}
\end{center}
\end{table}

\subsection{The CP-invariant case}

Motivated by the form of the results of Section~2 on two species,
we first consider the CP-invariant case,
which can be anticipated to be simpler than the generic one.

As far as statics is concerned, C-invariance implies
\beq
J_A=J_B=J,\qquad J_C=J_0.
\label{jcp}
\eeq
The Hamiltonian~(\ref{hambeg})
becomes the usual Blume-Emery-Griffiths Hamiltonian~\cite{beg}
\beq
\H=E_0-\sum_n\Bigl[(J+2J_0)S_nS_{n+1}+J\Bigr]S_nS_{n+1}.
\label{hambegc}
\eeq
As far as dynamics is concerned, CP-invariance yields
12 equalities among the 27 exchange rates:
\beq
\matrix{
\w_{AA}=\w_{BB},\quad\hfill
&\w_{AC}=\w_{CB},\quad\hfill
&\w_{BC}=\w_{CA},\quad\hfill
&\x_{AA}=\y_{BB},\hfill\cr
\x_{AB}=\y_{AB},\hfill
&\x_{AC}=\y_{CB},\hfill
&\x_{BA}=\y_{BA},\hfill
&\x_{BB}=\y_{AA},\hfill\cr
\x_{BC}=\y_{CA},\hfill
&\x_{CA}=\y_{BC},\hfill
&\x_{CB}=\y_{AC},\hfill
&\x_{CC}=\y_{CC}.\hfill\cr
}
\label{3cpcd}
\eeq

The analysis follows the lines of Section~2.
The algebra is however far more cum\-ber\-some,
so that intermediate expressions are too lengthy to be reported here.
Calculations have been worked out with the help of the software MACSYMA.
We start from the expressions for the total rates
$W_\out(\C)$ and $W_\in(\C)$ for a generic configuration $\C$,
similar to~(\ref{wout}) and~(\ref{win}),
which can be read off from Table~\ref{3d}.
The difference $W_\out(\C)-W_\in(\C)$
is then recast in terms of products of the spin variables~$S_n$.
We thus obtain an expression similar to~(\ref{dw}),
involving 42 different correlations of two to eight spin variables.
One example of a correlation of two variables is $\mean{S_1S_2}$,
whereas there is a unique correlation of eight variables:
$\mean{S_1^2S_2^2S_3^2S_4^2}$.
Requiring that the coefficients of all these correlations vanish,
we thus obtain 42 linear (but not independent) relations of the form
\beq
R_1=\dots=R_{42}=0,
\label{r42}
\eeq
where the $R_i$ are linear combinations of the exchange rates,
similar to $R_1$ and $R_2$ given in~(\ref{r1r2}).
In the CP-invariant situation under study,
equations~(\ref{3cpcd}) and~(\ref{r42}) together
yield 26 independent linear relations among the 27 exchange rates,
and therefore leave a single free parameter,
which can be fixed by choosing the time unit.
For consistency with the case of two species, we set
\beq
\w_{AA}+\w_{BB}=1.
\label{3nor}
\eeq
This normalization condition uniquely determines all the exchange rates.

We have therefore shown that there is a single totally asymmetric
CP-in\-va\-riant dynamics for three species of interacting particles
with stationary-state meas\-ure~(\ref{boltz}),~(\ref{ham3}).
This uniquely determined dynamics can be viewed
as a non-trivial extension to three species of the result~(\ref{resascp}).
The explicit expressions of the exchange rates are given in Table~\ref{3cp}.
This dynamics obeys pairwise balance
(see equation~(\ref{pb3}) below for the general case).
However, at variance with the case of two species,
the rates are not of the heat-bath form~(\ref{hb}).

\begin{table}[ht]
\caption{Expressions of the rates of
the totally asymmetric CP-invariant dynamics for three species.
The label $I$ stands for any particle species ($I=A$, $B$,~$C$).}
\label{3cp}
\begin{center}
\begin{tabular}{|c|c||c|c||c|c|}
\hline
Rate&expression&Rate&expression&Rate&expression\\
\hline
\haut$\w_{AA}=\w_{BB}$&$\frad{1}{2}$&
$\w_{BA}$&$\frad{{\rm e}^{4J}}{{\rm e}^{4J}+1}$&
$\x_{IA}=\y_{BI}$&$\frad{{\rm e}^{2J}}{2({\rm e}^{4J}+1)}$\\
\haut$\w_{AB}$&$\frad{1}{{\rm e}^{4J}+1}$&
$\w_{BC}=\w_{CA}$&$\frad{{\rm e}^{2J}({\rm e}^{2J}+1)}{2({\rm e}^{4J}+1)}$&
$\x_{IB}=\y_{AI}$&$\frad{1}{2({\rm e}^{4J}+1)}$\\
\haut$\w_{AC}=\w_{CB}$&$\frad{{\rm e}^{2J}+1}{2({\rm e}^{4J}+1)}$&
$\w_{CC}$&$\frad{{\rm e}^{2J}}{{\rm e}^{4J}+1}$&
$\x_{IC}=\y_{CI}$&$\frad{{\rm e}^{2J+2J_0}}{2({\rm e}^{4J}+1)}$\\
\hline
\end{tabular}
\end{center}
\end{table}

\subsection{The general case}

We now turn to the general totally asymmetric dynamics.
We view $J_A$, $J_B$, and $J_C$ as three independent coupling constants,
and consider the 27 exchange rates entering Table~\ref{3d}
as being a priori all different from each other.

Following the above procedure,
and choosing time units according to~(\ref{3nor}),
we are left after some lengthy algebra with a two-parameter family of dynamics
with stationary-state measure~(\ref{boltz}),~(\ref{ham3}).
An explicit parametrization of the rates is given in Table~\ref{3gene},
where the parameters $\alpha$ and $\beta$ enter linearly, and with the notation
\beq
f=\frac{1}{2({\rm e}^{2J_A+2J_B}+1)}.
\label{fdef}
\eeq
The form of the CP-invariant rates given in Table~\ref{3cp} has been helpful
in working out this parametrization of the general case.

\begin{table}[ht]
\caption{Explicit parametrization of rates
of the most general totally asymmetric dynamics for three species.
The notation $f$ is defined in~(\ref{fdef}).}
\label{3gene}
\begin{center}
\begin{tabular}{|c|c||c|c|}
\hline
Rate&expression&Rate&expression\\
\hline
$\w_{AA}$&$(1-\al+(1+\al)\eab)f$&
$\x_{BC}$&$(1+\al+\be)\ebc f$\\
$\w_{AB}$&$2f$&
$\x_{CA}$&$((1+\al)\eb+\be)f$\\
$\w_{AC}$&$(1-\al+(1+\al)\eb)f$&
$\x_{CB}$&$(1+\al+\be)f$\\
$\w_{BA}$&$2\eab f$&
$\x_{CC}$&$((1+\al)\ebc+\be)f$\\
$\w_{BB}$&$(1+\al+(1-\al)\eab)f$&
$\y_{AA}$&$(1-\al-\be\eac)f$\\
$\w_{BC}$&$(1+\al+(1-\al)\ea)\eb f$&
$\y_{AB}$&$(1-\al-\be\ec)f$\\
$\w_{CA}$&$(1-\al+(1+\al)\eb)\ea f$&
$\y_{AC}$&$(1-\al-\be)f$\\
$\w_{CB}$&$(1+\al+(1-\al)\ea)f$&
$\y_{BA}$&$(1-\al-\be\ec)\ea f$\\
$\w_{CC}$&$((1-\al)\ea+(1+\al)\eb)f$&
$\y_{BB}$&$((1-\al)\ea-\be\ec)f$\\
$\x_{AA}$&$((1+\al)\eb+\be\ec)f$&
$\y_{BC}$&$((1-\al)\ea-\be)f$\\
$\x_{AB}$&$(1+\al+\be\ec)f$&
$\y_{CA}$&$(1-\al-\be)\eac f$\\
$\x_{AC}$&$((1+\al)\eb+\be)\ec f$&
$\y_{CB}$&$((1-\al)\ea-\be)\ec f$\\
$\x_{BA}$&$(1+\al+\be\ec)\eb f$&
$\y_{CC}$&$((1-\al)\eac-\be)f$\\
$\x_{BB}$&$(1+\al+\be\ebc)f$&&\\
\hline
\end{tabular}
\end{center}
\end{table}

\begin{figure}
\begin{center}
\includegraphics[angle=90,width=.5\linewidth]{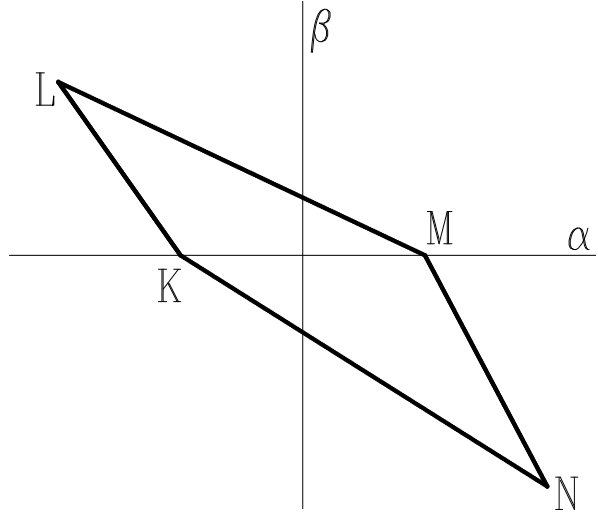}
\caption{\small
Typical shape of the quadrilateral domain $D$ of the $\al$-$\be$ plane
such that all the rates of Table~\ref{3gene} are positive (see text).}
\label{figquadra}
\end{center}
\end{figure}

The two parameters $\al$ and $\be$ run over some domain~$D$,
such that all the rates of Table~\ref{3gene} are positive.
It can be checked that~$D$ is an asymmetric quadrilateral,
shown schematically in Figure~\ref{figquadra}.
The co-ordinates of its vertices read
\beq
\matrix{
\haut\al_\K=-1,\quad\hfill&
\al_\L=-\frad{L_A+S_B}{L_A-S_B},\quad\hfill&
\al_\M=1,\quad\hfill&
\al_\N=\frad{L_B+S_A}{L_B-S_A},\hfill\cr
\haut\be_\K=0,\hfill&
\be_\L=\frad{2}{L_A-S_B},\hfill&
\be_\M=0,\hfill&
\be_\N=-\frad{2}{L_B-S_A},\hfill
}
\eeq
where $L_A$ (resp.~$S_A$) is the largest (resp.~the smallest)
of the three quantities
$\exp(2J_A+2J_C)$, $\exp(-2J_A+2J_C)$, and $\exp(-2J_A-2J_C)$,
and $L_B$ (resp.~$S_B$) is the largest (resp.~the smallest)
of the three quantities
$\exp(2J_B+2J_C)$, $\exp(-2J_B+2J_C)$, and $\exp(-2J_B-2J_C)$.

The general dynamics of Table~\ref{3gene} contains as special cases
several of the situations considered so far.
The CP-invariant situation corresponds to $\al=\be=0$.
The rates of Table~\ref{3cp} are thus recovered,
with the notations~(\ref{jcp}).
The case of two species is also recovered.
The rates~(\ref{resas}) are reproduced,
again with the notations~(\ref{jcp}), and with the identification
\beq
\delta=\frad{{\rm e}^{4J}-1}{{\rm e}^{4J}+1}\,\al.
\eeq

The family of dynamics described in Table~\ref{3gene}
obeys P-related pairwise balance.
This property is expressed by the nine relations
\beq
{\hskip -16pt}\matrix{
\w_{BA}=\eab\w_{AB},\quad\hfill
&\w_{CA}=\ea\w_{AC},\hfill
&\w_{BC}=\eb\w_{CB},\hfill\cr
\x_{BA}=\eb\x_{AB},\hfill
&\x_{AC}=\ec\x_{CA},\hfill
&\x_{BC}=\ebc\x_{CB},\hfill\cr
\y_{BA}=\ea\y_{AB},\hfill
&\y_{CA}=\eac\y_{AC},\quad\hfill
&\y_{CB}=\ec\y_{BC},\hfill\cr
}
\label{pb3}
\eeq
which are identically fulfilled by the rates of Table~\ref{3gene},
for any values of the parameters~$\al$ and $\be$.
In other words, the relations~(\ref{pb3})
are built in as a subset of~(\ref{r42}).

Pairwise balance has the following consequence.
Consider the partially asymmetric dynamics with uniform bias $p$,
where all the `right' moves, i.e, those of Table~\ref{3d},
take place with rates equal to $p$ times those given in Table~\ref{3d},
whereas the P-related `left' moves take place
with rates equal to $q$ times those given in Table~\ref{3d}, with $q=1-p$.
For example:
\beqa
&&AABC\to ABAC\quad\hbox{with rate}\ \ pw_{AC},\nonumber\\
&&CBAA\to CABA\quad\hbox{with rate}\ \ qw_{AC}.
\eeqa
For this uniformly biased dynamics, the total entrance and exit rates
for a given configuration read
\beqa
&&W_\in(\C)=pW^\rig_\in(\C)+qW^\lef_\in(\C),\nonumber\\
&&W_\out(\C)=pW^\rig_\out(\C)+qW^\lef_\out(\C),
\eeqa
with self-explanatory notations.
The stationarity condition for the totally asymmetric dynamics $(p=1)$
reads $W^\rig_\in(\C)=W^\rig_\out(\C)$.
On the other hand, P-related pairwise balance implies
$W^\rig_\in(\C)=W^\lef_\out(\C)$ and $W^\rig_\out(\C)=W^\lef_\in(\C)$.
We have therefore
\beq
W_\in(\C)=W^\rig_\in(\C)=W^\lef_\in(\C)
=W_\out(\C)=W^\rig_\out(\C)=W^\lef_\out(\C).
\eeq
These equations show that the partially asymmetric dynamics
with uniform bias $p$ has the same stationary-state measure
as the totally asymmetric one.
This dynamics interpolates between the symmetric (equilibrium) case ($p=1/2$)
and the totally asymmetric one ($p=1$).
The fact that the stationary-state measure is independent of the bias $p$
thus appears as a general consequence of P-related pairwise balance.

\section{Discussion}

In this paper we explicitly constructed classes of nonequilibrium dynamics
for two and three species of interacting particles,
i.e., asymmetric stochastic dynamics which do not obey detailed balance,
but whose nonequilibrium stationary-state measure is a prescribed measure.
We have chosen to work with finite-temperature canonical Gibbs measures
associated with spin Hamiltonians with nearest-neighbor interactions.
The stationary current, as well as many other observables in the stationary
state, can therefore be evaluated, at least in principle,
by means of the transfer-matrix formalism.
We have emphasized the role of the various symmetries
which can be imposed onto the dynamics.

For two species of interacting particles,
a situation first considered by KLS~\cite{kls},
stationary-state measures are associated with
the usual (anti)ferromagnetic Hamiltonian on the spin-1/2 Ising chain.
Our result for the most general dynamics is given by Table~\ref{2gene}.
This dynamics has five free parameters,
and does not obey pairwise balance in general.
Only the cases considered by KLS, namely the totally asymmetric dynamics
and the partially asymmetric dynamics with a uniform bias,
obey P-related pairwise balance,
where pairs of conjugate moves are related by parity.

We then turned to the novel situation of three species of interacting
particles.
Stationary-state measures are given by the most general Hamiltonian
involving pairs of neighboring particles of the same species.
This translates into a Blume-Emery-Griffiths spin-1 Hamiltonian.
We first restricted the search of dynamics
to the totally asymmetric case.
The most general situation is described by Table~\ref{3gene}.
This dynamics obeys pairwise balance.
It can therefore be extended to a partially asymmetric dynamics
with uniform bias $p$.
The three-parameter family of dynamics thus obtained
interpolates between the symmetric (equilibrium) case ($p=1/2$)
and the totally asymmetric one ($p=1$).

The most constrained class of stochastic dynamics
we have investigated is the CP-invariant totally asymmetric one.
For a prescribed stationary-state measure,
there is indeed a uniquely defined such dynamics,
with no free parameter, both for two species (see equation~(\ref{resascp}))
and for three species (see Table~\ref{3cp}) of interacting particles.

Throughout this work we have put a strong emphasis on the numbers
of free parameters in symmetric and asymmetric dynamics
leading to a given stationary-state measure.
Our results suggest the following rule:
asymmetric stochastic dynamics leading to a given nonequilibrium
stationary-state measure are far more constrained
than symmetric dynamics leading to the same measure as an equilibrium measure.

To close up, let us demonstrate that the above empirical rule
holds in a much broader class of stochastic models.
To do so, we have chosen to put the results of this work
in perspective with the following two characteristic examples,
which also belong to the realm of driven diffusive systems.

\medskip\noindent {\it Example~1.}

Our first example is much in the spirit of the present paper.
Consider a driven diffusive system
consisting of $K$ species of non-interacting particles on a ring,
denoted by $I=A,B,\dots$, where $K\ge2$ is arbitrary.

The most general exchange dynamics is defined by the $K(K-1)$ rates~$\u_{IJ}$
corresponding to the moves $IJ\to JI$ for $I\neq J$.
We look for dynamics such that the stationary-state measure is uniform,
i.e., all the configurations with given particle numbers~$N_I$ of each species
are equally probable.
This is indeed the right concept for a Gibbs measure
in the absence of interactions,
or, equivalently, in the limit of an infinite temperature.

The condition for having a uniform stationary-state measure reads
\beq
W_\out(\C)-W_\in(\C)=\sum_{IJ}\u_{IJ}(N_{IJ}-N_{JI})=0
\label{pdif}
\eeq
for every configuration $\C$.
The number of independent conditions on the rates imposed
by this equation can be evaluated as follows.
There are $K^2$ numbers of oriented pairs~$N_{IJ}$, which obey the sum rules
\beq
\sum_JN_{IJ}=\sum_JN_{JI}=N_I,\qquad\sum_IN_I=N.
\eeq
Only $(K-1)^2$ pair numbers are therefore linearly independent.
Equation~(\ref{pdif}) shown that each of the $(K-1)(K-2)/2$
antisymmetric combinations of these independent pair numbers
yields one condition.
The $K(K-1)$ rates therefore obey $(K-1)(K-2)/2$ conditions,
so that the general asymmetric exchange dynamics for $K$ species
of non-interacting particles depends on
\beq
A_K=\case{1}{2}(K-1)(K+2)
\eeq
dimensionful parameters.
On the other hand, for the symmetric exchange dynamics
obeying the detailed balance property $\u_{IJ}=\u_{JI}$,
the $K(K-1)/2$ rates are not constrained at all.
Indeed~(\ref{pdif}) vanishes identically.
The general symmetric exchange dynamics therefore depends on
\beq
S_K=\case{1}{2}K(K-1)
\eeq
parameters.
One has
\beq
A_K=S_K+K-1.
\label{dif}
\eeq

For two species ($K=2$), we have $A_2=2$ and $S_2=1$.
There is no condition on the exchange rates,
because there exists no antisymmetric combination of pair numbers.
Equation~(\ref{pairex}) indeed implies $N_{AB}=N_{BA}$.
As a consequence, the stationary-state measure
is uniform for any value of the rates $u_{AB}$ and $u_{BA}$.
Interpreting $A$ particles as particles and $B$ particles as holes,
we thus recover a known property of the ASEP~\cite{sch,der},
namely that its stationary-state measure is uniform, irrespective of the bias.

For three species ($K=3$), we have $A_3=5$ and $S_3=3$.
There is indeed one single antisymmetric combination of pair numbers:
\beq
Q=N_{AB}-N_{BA}=N_{BC}-N_{CB}=N_{CA}-N_{AC}.
\eeq
There is accordingly a single condition on the six exchange rates
for having a uniform stationary-state measure:
\beq
u_{AB}+u_{BC}+u_{CA}=u_{BA}+u_{CB}+u_{AC}.
\label{skew}
\eeq
This condition is known in the context of the matrix-product
formalism~\cite{ahr}.
It can be checked that~(\ref{skew}) is fulfilled by the rates
of Table~\ref{3gene} in the absence of interactions ($J_A=J_B=J_C=0$).
The only non-zero rates indeed read $u_{AB}=w_{IJ}=1/2$,
$u_{CB}=x_{IJ}=(1+\al+\be)/4$, and $u_{AC}=y_{IJ}=(1-\al-\be)/4$,
irrespective of $I,J$.
In the CP-invariant case, one has $u_{AB}=1/2$ and $u_{CB}=u_{AC}=1/4$.

Finally, for a large number of species ($K\gg1)$, asymmetric (driven,
nonequilibrium)
dynamics are far more constrained that symmetric (equilibrium) ones.
Indeed the condition of having a uniform stationary-state measure
roughly cuts off
half the parameters, reducing their number from $K(K-1)$ to $A_K\approx K^2/2$,
whereas for symmetric (equilibrium)
dynamics the $K(K-1)/2$ rates are not constrained.
The expression~(\ref{dif}) shows that the difference $A_K-S_K\approx K\ll S_K$
is relatively negligible for a large number of species.
In other words, for the uniform stationary-state measure,
the full space of nonequilibrium dynamics
is hardly larger than the subspace of equilibrium dynamics.

\medskip\noindent {\it Example~2.}

Our second example still belongs to the realm of driven diffusive systems,
albeit with multiple occupancies.
The results below strengthen our conclusion
and broaden its range of applicability.

Consider the class of dynamical urn models defined as follows.
$N$ particles are distributed among $M$ sites around a ring,
with multiple occupancies.
Let $N_m$ be the number of particles at site $m=1,\dots,M$.
The system is subjected to the following stochastic dynamics.

\noindent (i) a departure site $d$ is chosen uniformly at random.

\noindent (ii) a neighboring arrival site $a$ is chosen as
the right neighboring site ($a=d+1$) with probability~$p$,
or the left neighboring site ($a=d-1$) with probability $q=1-p$.

\noindent (iii) a particle is transferred from site $d$ to site $a$ at a rate
$W_{kl}$ which only depends on the occupancies $k=N_d$ and $l=N_a$
of the two sites.

The relevant question in the present context is the following one.
Under which conditions on the rates $W_{kl}$
is the stationary-state measure a product measure of the form
\beq
P(\C)=P(N_1,\dots,N_M)=\frac{1}{Z_{M,N}}\ p_{N_1}\dots p_{N_M}
\ \delta(N_1+\cdots+N_M,N)?
\label{fss}
\eeq
The answer to this question is known~\cite{cocozza} (see~\cite{lux}
for a simple presentation).
Consider first the case of an asymmetric dynamics ($p\ne1/2$).
The stationary-state measure is given by~(\ref{fss})
if and only if the rates $W_{kl}$ obey the two conditions
\beqa
&&p_{k+1}p_lW_{k+1,l}=p_kp_{l+1}W_{l+1,k},\label{m1}\\
&&W_{kl}-W_{k0}=W_{lk}-W_{l0}.\label{m2}
\eeqa
The first condition~(\ref{m1})
relates the rates $W_{kl}$ and the one-site factors $p_k$
of the stationary-state measure distribution.
The meaning of this relation is clear:
it just expresses P-related pairwise balance.
The second condition~(\ref{m2}), which does not involve the $p_k$,
is therefore more `kinematic' than `dynamical' in essence.

The zero-range process (ZRP) corresponds to the particular case
where the rates $W_{kl}=u_k$ only depend on the occupation of the departure
site.
The condition~(\ref{m2}) is then automatically satisfied,
whereas~(\ref{m1}) yields the following relation between the rates $u_k$
and the factors $p_k$:
\beq
u_k=\omega\;\frac{p_{k-1}}{p_k}
\eeq
for $k\ge1$, where the constant $\omega$ fixes the time unit.

The most general dynamical urn model with stationary-state product measure
is hardly more general than the ZRP.
Let us state the following result, skipping the proof.
For a given product measure of the form~(\ref{fss}),
with prescribed factors $p_k$,
the general solution of~(\ref{m1}) and~(\ref{m2})
is entirely determined by the one-dimensional array of rates $\al_k=W_{k0}$.
One has indeed (with $\al_0=0$)
\beq
W_{kl}=\frac{1}{p_kp_l}\sum_{m=0}^lp_{k+m}p_{l-m}(\al_{k+m}-\al_{l-m}).
\eeq
The rates $\al_k$ are the rates at which an empty site ($l=0$) is refilled,
by receiving one particle from a non-empty neighboring site
containing $k\ge1$ particles.

In the case of a symmetric dynamics ($p=1/2$),
only the first condition~(\ref{m1}) is requested~\cite{cocozza,lux}.
This relation expresses detailed balance.
The resulting stationary state is therefore an equilibrium state.
The condition~(\ref{m1}) determines the rates $W_{kl}$ for $k>l$
in terms of those for $k\le l$.

For a general dynamical urn model, the stationary product measure
thus depends on the one-dimensional array of rates $\al_k$
in the asymmetric case,
and on the two-dimensional array of rates $W_{kl}$ for $1\le k\le l$
in the symmetric case.

To sum up, the two above examples of driven diffusive systems
corroborate the picture which emerges from the results of the present work.
Asymmetric (driven, nonequilibrium) stochastic dynamics producing a given
stationary-state measure are far more constrained
(in terms of numbers of free parameters)
than symmetric dynamics producing the same measure as an equilibrium measure.

\end{document}